\documentclass[
reprint,
showpacs,
amsmath,amssymb,
aps,
pra,
floatfix
]{revtex4-2}

\usepackage{xcolor}
\usepackage{graphicx}
\usepackage{dcolumn}
\usepackage{bm}
\graphicspath{{Figures/}}
\usepackage{natbib}
\usepackage{hyperref}
\usepackage{amsmath}
\usepackage{comment}

\begin{document}
	
\title{Designing a compact cavity-enhanced source of entangled photons}

\author{\v{Z}iga Pu\v{s}avec}

\thanks{These two authors contributed equally}

\affiliation{Faculty for Mathematics and Physics, University of Ljubljana, Jadranska 19, 1000 Ljubljana, Slovenia}

\author{Lara Ul\v{c}akar}

\thanks{These two authors contributed equally}

\affiliation{Faculty for Mathematics and Physics, University of Ljubljana, Jadranska 19, 1000 Ljubljana, Slovenia}

\affiliation{Jozef Stefan Institute, Jamova 39, 1000 Ljubljana, Slovenia}

\author{Rainer Kaltenbaek}

\email{rainer.kaltenbaek@fmf.uni-lj.si}

\affiliation{Faculty for Mathematics and Physics, University of Ljubljana, Jadranska 19, 1000 Ljubljana, Slovenia}

\affiliation{Institute for Quantum Optics and Quantum Information, Austrian Academy of Sciences, Boltzmanngasse 3, 1090 Vienna, Austria}

\begin{abstract}
Entanglement will be the key resource of future large-scale quantum networks, enabling quantum communication and advanced quantum applications like distributed quantum sensing and distributed quantum computing. To this end, entanglement will have to be distributed over large distances and efficiently coupled to quantum devices at the network nodes. This requires the entangled photons to have wavelengths and bandwidths compatible with the quantum memories in quantum repeater nodes or quantum devices at client nodes. Here, we present a novel cavity-enhanced source design using two nonlinear crystals inside a single cavity. We provide detailed considerations balancing the complexity of the cavity design with the photon bandwidth and the entanglement quality.
\end{abstract}
\maketitle 

\section{Introduction}
Large-scale quantum networks will use entanglement as a resource for future quantum applications~\cite{Horodecki2009a, Wehner2018a}. While the distribution of entanglement has been demonstrated over a few hundred kilometers in optical fiber networks~\cite{Neumann2022,Zhuang2024a}, the distribution over larger distances requires to either use quantum repeaters~\cite{Briegel1998a} or to use space optical links~\cite{Yin2017a,Sidhu2021a,Krzic2023a}. Quantum repeaters typically rely on the efficient storage of the distributed entanglement in quantum memories~\cite{Azuma2023a}. This requires the matching between the photon bandwidth and the memory linewidth~\cite{Shinbrough2023a}. Depending on the type of quantum memory used, this can impose limitations on the bandwidth of the entangled photons to allow their efficient storage. While some quantum memories allow for high bandwidths up to many GHz~\cite{Zhang2023a}, others require a narrow bandwidth of a few MHz. For example, this is necessary to couple efficiently to the D2 transition of Rubidium~\cite{Schultz2008a}, or to long-lived optomechanical memories~\cite{Kristensen2024a}.

To effectively use the distributed entanglement, it must be coupled to quantum devices at client nodes of the network. This can pose additional requirements on the entanglement. For example, the bandwidth has to be sub-MHz for optomechanical devices with mechanical frequencies in the MHz regime~\cite{Kristensen2024a}. The bandwidth has to be even narrower if one wants to address a specific degree of motion of a trapped sub-micron particle. Such systems have, for example, become an attractive platform for realizing high-precision quantum sensors allowing for an unprecedented sensitivity in force measurements~\cite{Liang2023a}. Coupling entangled photons to the center-of-mass motion of such particles will require photon bandwidths below $100$~kHz~\cite{Delic2020a}.

Generating narrowband photons can be achieved using cavity-enhanced spontaneous parametric downconversion (SPDC)~\cite{Chuu2012,Lenhard2015a,Rambach2016a}. Other methods include spontaneous four-wave mixing in hot atomic vapor cells~\cite{Zhu2017, Hsu2021} or using laser-cooled atoms \cite{Wang2022, Bruns2022}. Recent works showed the generation of photon pairs compatible with the storage in atom-based~\cite{Tsai2018,Prakash2019,He2023} and solid-state~\cite{Liu2020,Onozawa2022} quantum memories. The performance of quantum memories was tested on weak coherent pulses ~\cite{Mannami2021,  Ito2023} and, in some cases, the actual storage of single photons has already been achieved~\cite{Buser2022,Davidson2023,Wei2024}.

This shows there has already been significant progress in producing and storing narrowband biphotons. Generating entanglement between two photons additionally requires making two biphoton processes indistinguishable. For example, this can be achieved by exciting indistinguishable processes in a single SPDC crystal~\cite{Kwiat1995a}, or by creating a superposition of two-photon amplitudes originating in multiple crystals~\cite{Kwiat1999a,Ljunggren2005}. 
Alternatively, the amplitudes can originate from multiple branches in an interferometer~\cite{Fedrizzi2007a}. Early works showed the generation of entangled cavity-enhanced photons using two crystals inside one cavity~\cite{Wang2004a} or by post-selecting an entangled state at the outputs of an interferometer~\cite{Bao2008a,Zhang2011a}. All these cases required significant spectral filtering to achieve narrowband entanglement. Rapid progress has been made in recent years in  simultaneously achieving entanglement and a narrow bandwidth of the generated photons. In Ref.~\cite{Gao2024}, the light from two cavity-enhanced SPDC crystals was used to generate entanglement with sub-GHz bandwidth in an interferometric setup. Refs.~\cite{Niizeki2020,Ito2023} showed the generation of narrowband entanglement using a frequency comb. These photons then were frequency converted to a wavelength compatible with solid-state quantum memories. The authors of Ref.~\cite{Arenskoetter2023} used a single crystal in a bi-directionally pumped bow-tie cavity to generate entangled photons compatible with a $^{40}\mathrm{Ca}^{+}$ quantum memory. 

In this paper,  we present a novel design for a cavity-enhanced SPDC source. The entangled photons are generated inside a single cavity. While this will require  a simultaneous resonance of four cavity modes, we will show that this is feasible with realistic experimental parameters. Even with a compact and simple cavity design, this will allow MHz linewidths with only moderate additional filtering. Significantly narrower bandwidths are possible if one allows for a more complex design. Additional dispersion compensation can be used to avoid the need for spectral filtering. A central goal of the paper is to investigate under which conditions realistic experimental parameters allow achieving high-quality entanglement. In Section~\ref{sec:source} we present how entanglement can be generated in a two-crystal setup using SPDC. In Section~\ref{sec:cavitySPDC}, we introduce the conditions for achieving simultaneous cavity resonances. We will apply these considerations in Section~\ref{sec:comparison} to two specific cases of phase matching. In Section~\ref{sec:cavity}, we determine the parameters necessary for the SPDC crystals and the cavity in order to achieve a specific bandwidth. Finally, in Section~\ref{sec:control} we discuss which experimental parameters can be tuned to find and to lock the necessary cavity resonances.

\section{A two-Crystal Source of Entanglement}
\label{sec:source}

\subsection{Spontaneous parametric downconversion}
SPDC is a second-order nonlinear optical process in which a high-energy pump photon with frequency $\omega_p$ is annihilated and generates a pair of lower-energy photons with frequencies $\omega_s$ and $\omega_i$~\cite{Couteau2018a}. These are denoted as the signal  and the idler photons. Energy conservation results in $\hbar\omega_p=\hbar\omega_s+\hbar\omega_i$. The wavevectors of the photons have to fulfill phase-matching (PM) conditions for the process to occur with significant probability. In a periodically poled crystal, the PM conditions can be modified by the poling period of the crystal to achieve quasi-phasematching (QPM). Depending on the photon polarizations, the crystal type and its orientation, different PM configurations are possible. In type-I PM, the signal and idler photons have the same polarization, which is orthogonal to the polarization of the pump. In type-0 PM, all three photons have the same polarization. In type-II PM, the signal and idler photons have orthogonal polarizations.

The bandwidth of a single-pass SPDC process is determined by the phase mismatch $\Delta k=k_s+k_i-k_p-2\pi/\Lambda$. Here, $\Lambda$ is the period of the periodic poling of the SPDC crystal. The amplitude of the photons generated in a crystal of length $l$ is proportional to $(\sin\xi/\xi)^2$ with $\xi=\Delta k l/2$. The FWHM bandwidth $\Delta \nu_\text{SPDC}$ of the SPDC photons can then be determined from  $\xi_{\text{HWHM}}\approx 1.39$. Expanding $\Delta k$ up to linear order in signal frequency and temperature yields:
\begin{equation}
    \Delta k =\left( \frac{\partial \Delta k}{\partial \omega_s}\right)_{\omega_p}\Delta\omega_s +\frac{\partial \Delta k}{\partial T}\Delta T = \frac{2\xi_{\text{HWHM}}}{l}.
    \label{eq:FWHMspdc}
\end{equation}
Here, $T$ is the crystal temperature, which affects the refractive index and the poling of the nonlinear crystal~\cite{Jundt1997, Gayer2008}. The SPDC bandwidth $2 \Delta\omega_s$ can be determined from the solution of Eq.~\ref{eq:FWHMspdc}. In a standing-wave cavity, the length $l$ in that equation has to be replaced by twice the crystal length because the pump will pass each crystal twice in one round trip (see Fig.~\ref{fig:setup}).

\subsection{Generating Entanglement}

\begin{figure}
    \centering
    \includegraphics[width=0.4\textwidth]{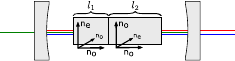}
    \caption{Schematic of nonlinear crystals of lengths $l_{1}$ and $l_{2}$ in a Fabry Per\'{o}t cavity. The green line indicates the pump. Blue and red lines indicate the signal and idler photons generated by SPDC. The optical axes of the crystals are rotated $90^{\circ}$.}
    \label{fig:setup}
\end{figure}

The source (Fig.~\ref{fig:setup}) is based on a single cavity that contains two nonlinear crystals rotated by 90 degrees relative to each other, inspired by \cite{Ljunggren2006, Trojek2008, Chuu2011, Chuu2012}. In principle, this allows for a very compact cavity design if the crystal properties are chosen according to the considerations we will present here. The crystals can be of different lengths, $l_1$ and $l_2$ such that $l=l_1+l_2$. We pump the crystals with continuous-wave light at frequency $\omega_p$ and with the polarization chosen to balance the brightness of both SPDC processes.
Depending on its polarization, the pump will create a pair of SPDC photons in the first or the second crystal.  Like the crystal axes, the polarization directions of the photons created in the first crystal will be perpendicular to those created in the second crystal.

Since the pump photon can generate an SPDC pair either in the first or in the second crystal, the corresponding two-photon state will be a linear combination of these two processes:
\begin{equation}
    \vert\psi\rangle=\frac{1}{\sqrt{2}}\Big( a_{1}^\dagger(\omega_s)a_{2}^\dagger(\omega_i)+e^{i\varphi}\,a_{3}^\dagger(\omega_i)a_{4}^\dagger(\omega_s)\Big)\vert\mathrm{vac}\rangle.
    \label{eq:state}
\end{equation}
$\varphi$ is a relative phase and $a^\dagger_{j}(\omega)$ is a boson creation operator of mode $j$ at frequency $\omega$. $j$ can indicate the polarization or the crystal of origin. 
The polarization of the photons generated is determined by the PM condition. For type-II SPDC, the state will be a superposition of the Bell states $\vert\psi^{\pm}\rangle$, while for type-0 and type-I SPDC it will be a superposition of Bell states $\vert\Phi^{\pm}\rangle$~\footnote{ $\vert\psi^{\pm}\rangle = 
(1/\sqrt{2})\left(\vert H V\rangle \pm \vert V H\rangle\right)$ and $\vert\Phi^{\pm}\rangle = 
(1/\sqrt{2})\left(\vert H H\rangle \pm \vert V V\rangle\right)$ are the four Bell states.}. The feasibility of generating entanglement depends on the frequencies and polarizations of generated photons. If the signal and idler photons originate from type-II SPDC and have degenerate frequencies, the amplitudes from different crystals will result in a product state $a^{\dagger}_H(\omega)a^{\dagger}_V(\omega) (1+e^{i\varphi})\vert\mathrm{vac}\rangle$.
Using our source design with type-II PM to generate entanglement is therefore not possible for degenerate frequencies. 
For type-0 and type-I SPDC, one cannot spatially separate the photons if the frequencies are degenerate. For these reasons, we will focus on two specific cases: (1) the photons produced have \textit{near}-degenerate wavelengths suitable for the low-loss distribution in telecommunication fibers, (2) only one photon is at a wavelength appropriate for telecommunication fibers, while the second photon's wavelength is chosen to allow for efficiently coupling to a specific type of quantum memory.

\section{Cavity-enhancement}
\label{sec:cavitySPDC}
\subsection{Simultaneous resonances}
Often, the bandwidth of SPDC photons is much broader than the acceptance bandwidth of quantum memories~\cite{Wang2019a,Heller2022a}. The effective bandwidth can be reduced by coupling the process to an optical cavity \cite{Boyd1966, Boyd1968, Chuu2011, Chuu2012, Tsai2018}. Only the portions of the photon spectra that are resonant with the narrow cavity modes will be enhanced. In comparison, the remainder of the spectra will be strongly suppressed. We describe these cavity modes
by defining a mode number $m_j$ for each mode $j$. It represents the phase shift that light in mode $j$ acquires during one cavity round trip. For a two-crystal type-II source, the four corresponding mode numbers are:
\begin{equation}
    \begin{split}
            m_{1} (\omega_s) & =\frac{\omega_s}{\pi c} \left( l_{1}n_{o}(\omega_s) +  l_{2}n_{e}(\omega_s) + l_{\text{air}} \right), \\
            m_{2} (\omega_i)  & =\frac{\omega_i}{\pi c} \left( l_{1}n_{e}(\omega_i) +  l_{2}n_{o}(\omega_i) + l_{\text{air}} \right), \\
            m_{3} (\omega_s) & =\frac{\omega_s}{\pi c} \left( l_{1}n_{e}(\omega_s) +  l_{2}n_{o}(\omega_s) + l_{\text{air}} \right), \\
            m_{4} (\omega_i)  & =\frac{\omega_i}{\pi c} \left( l_{1}n_{o}(\omega_i) +  l_{2}n_{e}(\omega_i) + l_{\text{air}} \right).
    \end{split}    
\end{equation}
Here, $m_{1}$ and $m_{2}$ represent the downconverted signal and idler mode numbers originating in the first crystal, while $m_{3}$ and $m_{4}$ represent the equivalent mode numbers for the second crystal. 
For a type-0 or type-I source, the corresponding mode numbers are:
\begin{equation}
    \begin{split}
            m_{1} (\omega_s) & =\frac{\omega_s}{\pi c} \left( l_{1}n_{o}(\omega_s) +  l_{2}n_{e}(\omega_s) + l_{\text{air}} \right), \\
            m_{2} (\omega_i)  & =\frac{\omega_i}{\pi c} \left( l_{1}n_{o}(\omega_i) +  l_{2}n_{e}(\omega_i) + l_{\text{air}} \right), \\
            m_{3} (\omega_s) & =\frac{\omega_s}{\pi c} \left( l_{1}n_{e}(\omega_s) +  l_{2}n_{o}(\omega_s) + l_{\text{air}} \right), \\
            m_{4} (\omega_i)  & =\frac{\omega_i}{\pi c} \left( l_{1}n_{e}(\omega_i) +  l_{2}n_{o}(\omega_i) + l_{\text{air}} \right).
    \end{split}    
\end{equation}
A single mode $j$ is resonant with the cavity if its frequency is an integer multiple of the free spectral range $\text{FSR}_j = \left\vert\partial_\omega m_j \right\vert^{-1} /2\pi$. Here, $\partial_{\omega}$ denotes the partial derivative with respect to $\omega$. To achieve narrowband entanglement between photons in these four modes, all of them have to be resonant with the cavity simultaneously. We denote this as a quadruple resonance.

\begin{figure}
    \centering
    \includegraphics[width=0.5\textwidth]{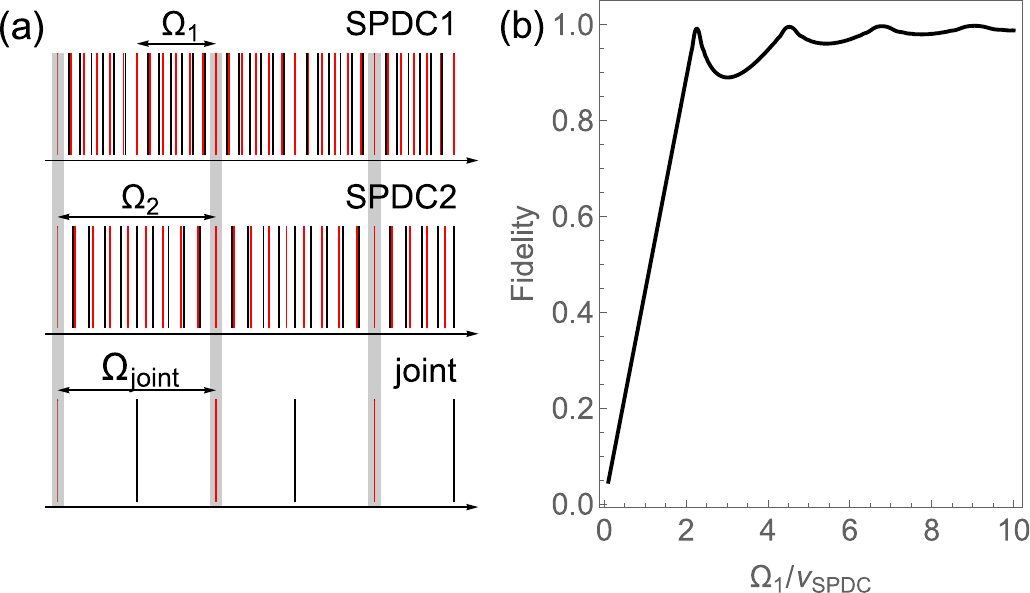}
    \caption{(a) Illustration of the cluster spacings of double resonances and the joint cluster spacing for quadruple resonances.    
    Red and black lines indicate idler and signal resonances, respectively. Top segment: occurrence of double resonances for SPDC in the first crystal. Middle segment: double resonances in the second crystal. Bottom segment: double resonances for both crystals, and the gray-shaded areas highlight quadruple resonances, where the signal and idler modes from both SPDC processes are resonant simultaneously. 
    (b) The fidelity of the state of a photon pair from a single crystal in well-defined frequency modes as a function of the cluster spacing normalized to the SPDC bandwidth.}
    \label{fig:spectrum}
\end{figure}

Due to the frequency-dependent birefringence of the crystals, and due to their different lengths, the FSRs of the individual downconverted modes will be different. Simultaneous resonances of multiple modes will occur only over specific frequency spacings much greater than individual FSRs. This is known as cluster spacing. The cluster spacing $\Omega_1$ of a double resonance between modes $1$ and $2$ can be expressed in terms of their FSRs  \cite{Chuu2011, Chuu2012}:
\begin{equation}
    \Omega_1 = 
    \frac{\textrm{FSR}_{1} \textrm{FSR}_{2}}{\vert\textrm{FSR}_{1}-\textrm{FSR}_{2}\vert}.
    \label{eq:FSRcluster}
\end{equation}
Similarly, the cluster spacing $\Omega_2$ is calculated for the modes $3$ and $4$.  For quadruple resonances, we define the joint
cluster spacing $\Omega_{\textrm{joint}}$ by replacing  $\textrm{FSR}_1$ and $\textrm{FSR}_2$ in Eq.~\ref{eq:FSRcluster} with $\Omega_1$ and $\Omega_2$, respectively. Fig.~\ref{fig:spectrum}(a) illustrates this relationship.

One wants the joint cluster spacing sufficiently small such that finding a quadruple resonance is achievable by tuning control parameters within a range that is experimentally feasible. An additional challenge is finding the right ratio between the cluster spacing and the SPDC bandwidth. 
If the cluster spacing in an individual crystal is smaller than the corresponding SPDC bandwidth, multiple cavity modes at different frequencies will be enhanced at the same time. The presence of multiple enhanced frequency modes will reduce the fidelity with the targeted two-photon state. 

\subsection{The state fidelity}
\label{app:fidelity}

In order to quantify the quality of the state produced in an individual crystal, we calculate its fidelity with the biphoton state we target: $a_{1}^\dagger(\omega_s)a_{2}^\dagger(\omega_{i})\vert\mathrm{vac}\rangle$.
The output state of the cavity-enhanced SPDC in crystal $\alpha$ will be a linear combination of all possible frequency combinations of signal and idler photons within the SPDC bandwidth: $\sum_j A^{\alpha}_j a_{1}^\dagger(\omega_s+j\Omega_{\alpha})a_{2}^\dagger(\omega_i-j\Omega_{\alpha})\vert\mathrm{vac}\rangle$. Here, $A^{\alpha}_j=\mathrm{sinc}(l_{\alpha}\Delta k_{j}/2)$ are normalized amplitudes depending on the phase mismatch of the SPDC process $\Delta k_{j,\alpha}$,
\begin{equation}
    \begin{split}
        \Delta k_{j,\alpha}\approx\frac{j\Omega_{\alpha}}{c}\left(-n_{o/e}(\omega_s)-\omega_s\Big(\frac{\partial n_{o/e}}{\partial \omega}\right)_{\omega_s}+\\
        n_{e/o}(\omega_i)+\omega_i\left(\frac{\partial n_{e/o}}{\partial \omega}\right)_{\omega_i}\Big).
        \label{eq:dk}
    \end{split}
\end{equation}
$\alpha=1,2$ denotes crystal $1$ or $2$. Fig.~\ref{fig:spectrum}(b) shows the calculated fidelity of the output state of crystal 1 for type-II PM with a product state $a_{H}^\dagger(\omega_s)a_{V}^\dagger(\omega_i)\vert\mathrm{vac}\rangle$ as a function of cluster spacing normalized to the SPDC bandwidth. A fidelity above 0.9 is achieved for
\begin{equation}
    \Omega/ \Delta\nu_{\text{SPDC}} > 2.
    \label{eq:ClusterBandwidth}
\end{equation}
A source fulfilling this bound for both crystals in the cavity will not require spectral filtering because only the processes at two frequencies with narrow bandwidths will be enhanced.

\section{Comparing resonance spacings to SPDC bandwidth}
\label{sec:comparison}
In the following, we will assume realistic experimental parameters and analyze the feasibility of the source design for the cases of near-degenerate and highly non-degenerate frequencies using type-0 and type-II PM. The near-degenerate source is centered at $1550$ nm and is suitable for distribution of entangled pairs through optical fibers. For the highly non-degenerate source, we consider entangled photons at $1550$ nm and $780$ nm - the former for long-distance communication and for potentially coupling to optomechanical devices, the latter for coupling to Rubidium quantum memories.
For the SPDC crystals, we assume $5\%\,$MgO-doped periodically poled Lithium Niobate (PPLN) due to its high nonlinearity and transparency at $1550$~nm~\cite{Gayer2008, Jundt1997}. 

Our analysis starts by finding the PM condition for the chosen wavelengths for an experimentally feasible periodic poling and temperature of the crystal to find the lowest possible SPDC bandwidth. For these parameters, we calculate the cluster spacing and the SPDC bandwidth and evaluate them in relation to Eq.~\eqref{eq:ClusterBandwidth}. Based on this, we fix the type of PM and calculate the joint cluster spacing. We find the ratio of crystal lengths that gives a reasonably low joint cluster spacing such that quadruple resonance can be achieved by tuning the experimental parameters.
A similar analysis can be performed for entangled photons at other combinations of frequencies. 

\subsection{Near-degenerate source}
First, let us consider a near-degenerate source suitable for distributing entanglement over telecom fibres. We assume a pump wavelength of $775\,$nm and near-degenerate signal and idler modes at $1550\,\textrm{nm} \pm\Delta \lambda/2$. $\Delta\lambda$ corresponds to a small frequency shift $\Delta\omega = 2\pi\times 80\,$MHz. The shift is chosen such that a typical AOM can shift the frequency of one of the photons, potentialy making the source degenerate at detection stage.
\begin{figure}[h]
    \centering
    \includegraphics[width=0.23\textwidth]{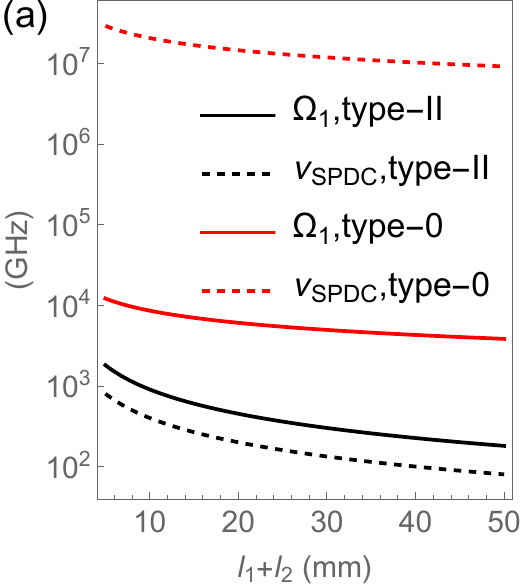} \includegraphics[width=0.23\textwidth]{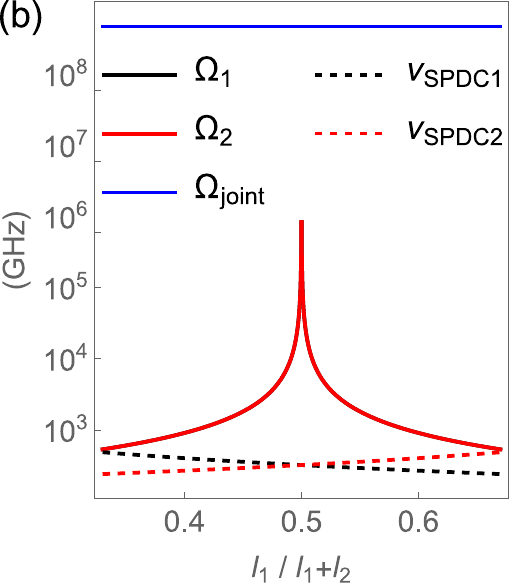}
    \caption{Cluster spacings vs SPDC bandwidths in the C-band at near-degenerate wavelengths 
    centered around $1550\,$nm with a frequency difference of $80\,$MHz. 
    (a) Comparison of type-II and type-0 phase matching (PM) when creating photon pairs in a single crystal. The dashed lines show the bandwidths of the SPDC processes, the solid lines show the corresponding cluster spacings
    as a function of the joint crystal length $l_1+l_2$ at a fixed ratio $l_1/(l_1+l_2)=0.4$. Because the SPDC bandwidth is larger than the cluster spacing, type-0 PM does not allow generating photon pairs of well-defined frequencies.
    (b) Comparing the joint cluster spacing to the individual cluster spacings and the SPDC bandwidths in the two crystals for type-II PM. We plot them as a function of the ratio of crystal lengths for a fixed joint crystal length $l_1+l_2=10\,$mm. The joint cluster spacing for type-II PM is too large to allow finding a quadruple resonance under realistic conditions.\label{fig:ClusterNearDeg}
    }
\end{figure}
Fig.~\ref{fig:ClusterNearDeg}(a) shows the SPDC bandwidth and the cluster spacing of the modes originating in crystal 1 for type-II and type-0 PM, as a function of the joint crystal length~\footnote{Type-0 cluster spacing was calculated using a second order expansion. See Supplemental Material~\cite{SI} and ~\cite{Eckardt1991} for more details. For the SPDC photons originating in crystal 2, identical considerations can be made.}.
For type-0 SPDC, the bandwidth and the cluster spacing are exceedingly high, whereas for type-II SPDC, the SPDC bandwidth is notably less than the cluster spacing. This means that a source based on type-0 PM is not possible but a setup based on type-II may be feasible. In particular, using type-II PM can ensure that cavity enhancement is achieved for the intended single pair of frequencies with high fidelity. 

Generating entangled pairs, however, requires a quadruple resonance, and we will see in the following that this is not possible for a near-degenerate source even for type-II PM. Fig.~\ref{fig:ClusterNearDeg}(b) shows that the cluster spacings of the double resonances grow significantly as the ratio of $l_1/(l_1+l_2)$ approaches $0.5$. At this point, the four crystal modes nearly become degenerate for small $\Delta\omega$. One can see that the joint cluster spacing is extremely large for all length ratios. Two quadruple resonances would be displaced by a wavelength shift of more than $1~\mu$m. A type-II near-degenerate source could therefore produce narrowband photon pairs, but an entangled state cannot be achieved since one cannot find a quadruple resonance in practice. 

\subsection{Highly non-degenerate source}
Secondly, let us consider the case of a highly non-degenerate source. We assume a pump wavelength of $519\,$nm to generate signal photons at the Rubidium transition $\lambda_s=780.24\,$nm and idler photons at $\lambda_i=1550\,$nm. The idler photons would then be suitable for a long-distance distribution over telecom fibers. If they are sufficiently narrowband, one could also couple the idler photons to optomechanical systems.
\begin{figure}[h]
    \centering
    \includegraphics[width=0.23\textwidth]{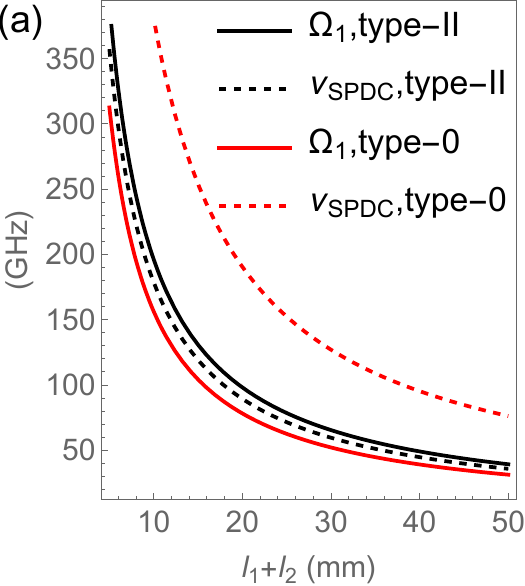} \includegraphics[width=0.23\textwidth]{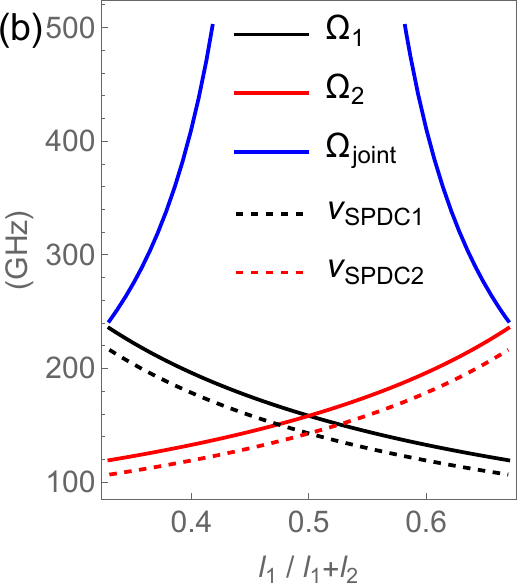}\\
    \includegraphics[width=0.23\textwidth]{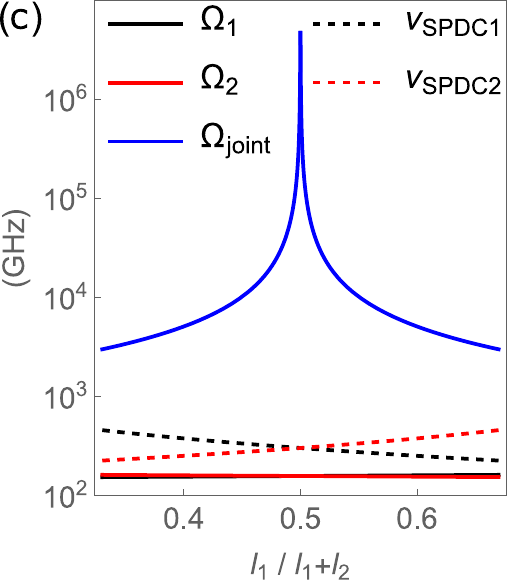} \includegraphics[width=0.23\textwidth]{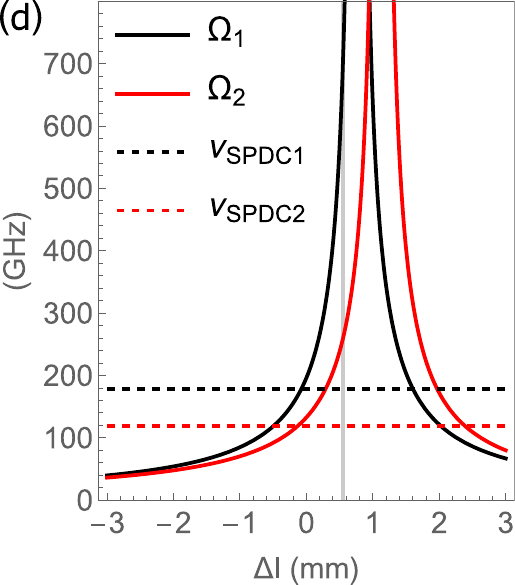}
    \caption{Cluster spacings vs SPDC bandwidths for highly non-degenerate wavelengths $\lambda_i=780.24\,$nm and $\lambda_s=1550\,$nm.
    (a) Comparison of type-II and type-0 phase matching (PM) when creating photon pairs in a single crystal. The dashed lines show the bandwidths of the SPDC processes, the solid lines show the corresponding cluster spacings
    as a function of the joint crystal length $l_1+l_2$ at a fixed ratio $l_1/(l_1+l_2)=0.4$.    
    (b) Comparing the joint cluster spacing to the cluster spacings and the SPDC bandwidths in individual crystals for type-II PM. We plot them as a function of the ratio of crystal lengths for a fixed joint crystal length $l_1+l_2=10\,$mm. 
    (c) Comparing the joint cluster spacing to the cluster spacings and the SPDC bandwidths in individual crystals for type-0 PM. For all length ratios, the joint cluster spacing is larger than $1\,$THz, which makes the quadruple resonance essentially impossible to find. (d)  Cluster spacing and SPDC bandwidth of a type-II source. An additional optical-path delay $\Delta l$ is added to modes 2 and 4.
    }
    \label{fig:ClusterNondeg}
\end{figure}

Fig.~\ref{fig:ClusterNondeg}(a) shows the SPDC bandwidth and the cluster spacing of a double resonance in crystal 1 as a function of the joint crystal length. Note that both the cluster spacing and the SPDC bandwidth reduce as the crystal length increases. For both types of phase matching, the SPDC bandwidth is too large to generate photon pairs where the frequency of each individual photon is well defined. 
The corresponding fidelity will therefore be only $0.5$ for a type-II source and $0.2$ for a type-0 source (see Fig.~\ref{fig:spectrum}(b)). Type-0 PM has the additional disadvantage that the joint cluster spacing is on the order of THz, as shown in Fig.~\ref{fig:ClusterNondeg}(c). This makes it impossible to find the quadruple resonances necessary for generating entangled photon pairs.

We will therefore focus on the more favorable type-II PM. To achieve that, we investigate the dependence of the joint cluster spacing on the ratio of the crystal lengths as shown in Fig.~\ref{fig:ClusterNondeg}(b). If the crystal lengths are the same, the joint cluster spacing will diverge since the two pairs of cavity modes become identical and will have the same cluster spacing for double resonances. To allow finding a quadruple resonance to generate entangled states, we therefore choose crystals of differing lengths. Fig.~\ref{fig:ClusterNondeg}(b) shows that a length ratio between 0.3 and 0.4 provides reasonable values for the joint cluster spacing. In this range, it is still sufficiently small to allow finding a quadruple resonance for realistic experimental parameters. Even then, the SPDC processes in the individual crystals can still produce photons in multiple frequency modes. 

We suggest two solutions to ensure a high quality of the entanglement generated.  One is to use a compact cavity design in combination with moderate spectral filtering. The practicality of this solution will depend on the spacing of the double resonances. The second solution is to introduce a relative optical path delay 
between the signal and idler modes to increase the individual cluster spacings. While this will result in a more complex and larger cavity design, it will allow generating narrowband entangled photons without the need for any additional spectral filtering. If the complexity and scale of the source do not pose a limitation, larger cavity lengths can allow even significantly narrower linewidths. In particular, we introduce an optical path delay $\Delta l_2$ to idler mode $2$ and $\Delta l_4$ to mode $4$. We model this by adding the term $\Delta l_{2/4}\,\omega_i/\pi c$ to the mode numbers $m_{2/4}$. This will leave the SPDC bandwidths and the joint cluster spacing unchanged, but it will affect the cluster spacings for the SPDC processes in the individual crystals. The comparison of cluster spacings and SPDC bandwidths as a function of path delay is shown in Fig.~\ref{fig:ClusterNondeg}(d). For simplicity, we take $\Delta l_2=\Delta l_4=\Delta l$. For $\Delta l=550\,\mu$m, the cluster spacings of the double resonances from both SPDC processes become twice as large as the corresponding SPDC bandwidths. For these values, one can achieve a fidelity of $0.93$ with the targeted maximally polarization entangled state $\vert\psi^{-}\rangle$.

\section{Cavity parameters}
\label{sec:cavity}

The bandwidth of the photons leaving the cavity is determined by the linewidth of the cavity mode, which in turn will depend the intracavity losses and the cavity length. These are the parameters one can use to optimize the cavity characteristics and define the bandwidth of the photons generated. In the following, we will discuss bandwidths achievable using realistic experimental parameters. For our calculation, we assume the nonlinear crystals to be PPLN with a joint crystal length of $10\,$mm. 

\begin{figure}
    \centering
    \includegraphics[width=0.4\textwidth]{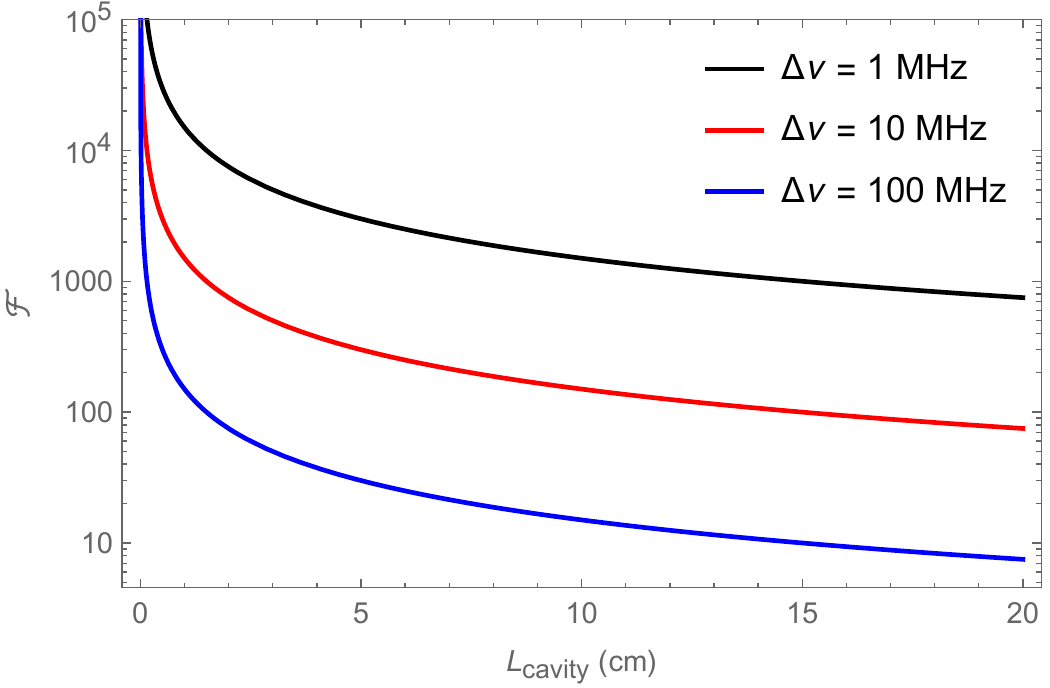}
    \caption{The finesse required to achieve a specific cavity linewidth as a function of the cavity length. In practice the desire for increased cavity length needs to be balanced with a need for a stable and compact cavity design.
    \label{fig:lossesfinesse}}
\end{figure}
The FWHM linewidth $\Delta \nu$ of the cavity is related to the finesse, which is defined as:
\begin{equation}
     \mathcal{F}  = \frac{\textrm{FSR}}{\Delta \nu},
     \label{eq:finesse_def}
\end{equation}
where $\textrm{FSR} = c / 2 L_{\textrm{cavity}}$ for a standing-wave resonator. The dependence of the finesse on the cavity length $L_{\textrm{cavity}}$ and its relation to the cavity linewidth are shown in Fig.~\ref{fig:lossesfinesse}. While increasing the cavity length will allow narrower linewidths, this will negatively impact the stability and compactness of our source of entangled photons. 

If we assume a simple Fabry-P\'{e}rot cavity with two mirrors of an identical radius of curvature, the optimal cavity length can be determined as follows. One first finds the optimal focusing condition \cite{Boyd1966,Boyd1968} and then matches the curvature of the optimally focused beam to the curvatures of the cavity mirrors. If one assumes the radii of curvature to be $15\,$mm, and if one takes into account the refractive index of PPLN, one arrives at an effective cavity length of $53\,$mm.

Once the cavity length is determined, one can aim to decrease the cavity linewidth by choosing appropriate mirror reflectivities and by minimizing the intra-cavity losses. In particular, the finesse depends on the mirror reflectivities $R_{1}$ and $R_{2}$ and the intracavity losses $\eta$ in the following way:
\begin{equation}
    \mathcal{F}  = \pi \arccos \left( \frac{4 \sqrt{R_{1} R_{2}(1-\eta)}-R_{1} R_{2} (1-\eta)-1}{2 \sqrt{R_{1} R_{2}(1-\eta)}} \right)^{-1}.
    \label{eq:finesse_losses}
\end{equation}
A derivation of an equivalent expression for the finesse can be found in Ref.~\cite{Ismail2016}.
Increasing the finesse can be achieved by decreasing the overall losses in the cavity through choosing high reflectivities for the cavity mirrors, high-quality antireflection coatings for the crystal end faces and low-absorption nonlinear crystals. 

\begin{figure}
    \centering
    \includegraphics[width=0.4\textwidth]{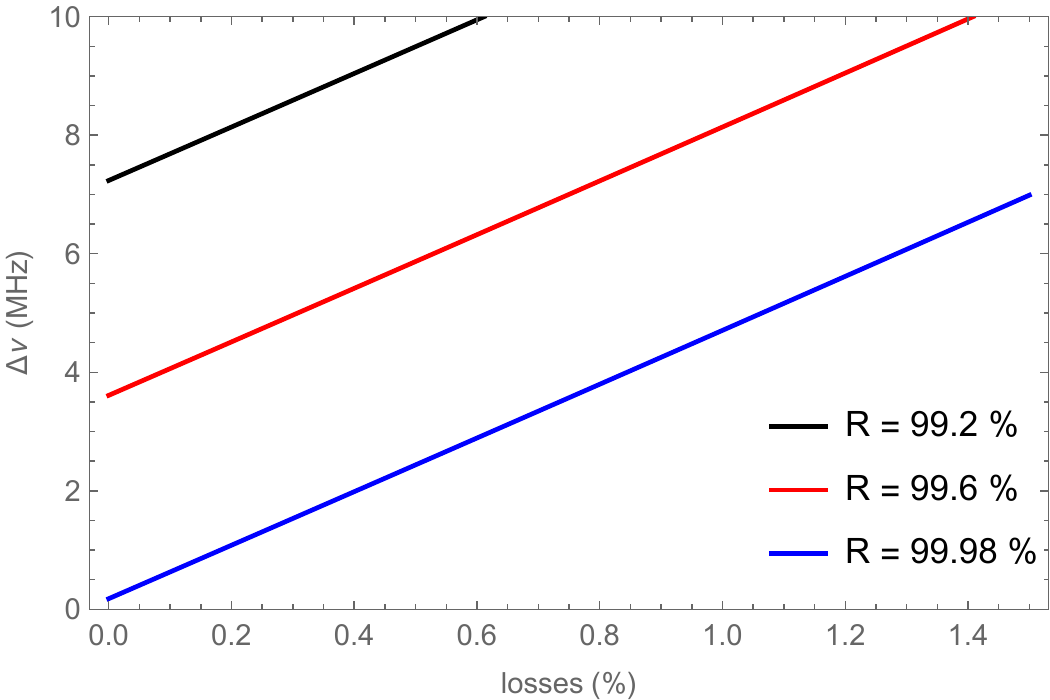}
    \caption{The dependence of the cavity linewidth on the intracavity losses for specific reflectivities of the cavity mirrors. The cavity length is $53\,$mm. By picking the desired cavity linewidth and the available mirror reflectivities one can determine the criteria for allowed losses in the cavity. Manufacturers usually only provide upper boundaries for reflectivities and antireflection coatings, which needs to be acknowledged when designing the cavity.
    \label{fig:losseslinewidth}}
\end{figure}

Let us now estimate an upper limit on the reflectivities of the antireflection coating for the wavelengths of the SPDC photons. To this end, we fix the two mirror reflectivities to be the same ($R_{1}=R_{2}\equiv R$) and then calculate the linewidth as a function of the corresponding intra-cavity losses using Eq.~\eqref{eq:finesse_def}. and Eq.~\eqref{eq:finesse_losses}.
In Fig.~\ref{fig:losseslinewidth}, this dependence is shown for a $1\,$cm long PPLN crystal and for several choices of mirror reflectivities. We neglected the losses in PPLN (less than  0.0005 $\mathrm{cm}^{-1}$ for considered wavelengths \cite{Schwesyg2011}) compared to the losses due to the anti-reflection (AR) coatings (for example, crystals with 0.2\% AR coating are available from manufacturer HC Photonics).

For example, if we assume $0.2\%$ AR coatings, the total intracavity losses will be about $0.8\%$ because we of the four air-to-crystal boundaries the photons will encounter per round-trip. This will result in a cavity linewidth of about $4\,$MHz for mirror reflectivities of $99.98\%$.

\section{Finding and stabilizing the cavity resonances}
\label{sec:control}

Maintaining resonances of the cavity requires active stabilization of the cavity. This can be done using a standard Pound-Drever-Hall technic \cite{Drever1983}. Because the four resonances are not independent due to energy conservation of SPDC, it will be sufficient to prevent the drift of two modes to stabilize the overall system. Stabilizing one mode prevents an absolute drift of the set of resonances, while stabilizing the second mode prevents relative drifts. The piezo on one of the cavity mirrors can follow the cavity drifts to maintain it on resonance with the reference pump light. The reference laser can be locked to the second mode to prevent the relative positions of resonances drifting apart. 

Locking can be done in two regimes. One option is to use a continuous lock. In this case, the cavity resonances are stabilized in reference to error signals of resonances that are one FSR detuned from the operating wavelengths. These strong laser signals can then be filtered from the single-photon modes using filtering cavities without affecting the entangled state. The suppression of light from the locking lasers can be enhanced by using a bow-tie cavity design, which allows an additional spatial distinction between counter-propagating modes~\cite{Rambach2016a}. Another way to lock is to operate the cavity alternatingly in two regimes \cite{Tsai2018}: (1) an optical parametric oscillator (OPO) regime for locking, where strong reference laser light above the OPO threshold pumps the crystals inside the cavity. (2) Periods of time during which entangled pairs are generated. In regime (2), no reference light is present, and the cavity is left to drift freely for a short time. This requires appropriately fast optical switches, intensity modulators and circulators to protect single photon detectors from strong reference laser light, and to switch the pump beam above and below the OPO threshold. 
 
The circulators will also play an essential role in allowing the efficient collection of single photons generated inside the cavity into single mode fibers. This coupling to a single mode will act as a spatial filter for light coupled into and out of the cavity. The amount of light collected will primarily depend on the proper impedance matching of the cavity. This will also strongly depend on the choice of using a reflective or a transmissive cavity design.

As presented in previous sections, an important challenge will be to find a suitable quadruple resonance. Finding and stabilizing the cavity resonances can be achieved by controlling several degrees of freedom: 1) the temperatures of the nonlinear crystals 2) the frequency of the pump light, and 3) moving a cavity mirror with a piezo.
For example, tuneable narrowline lasers typically offer several 10 GHz of mode-hop free frequency tuning. If one assumes a PPLN SPDC crystal of about $10\,$mm length, changing its temperature by $1\,$K will shift the corresponding cluster spacing for the signal and idler frequencies by about $2\,$THz. This large dependence on temperature means that the crystal temperature has to be stabilized with mK precision. Tuning the temperature will be essential for initially finding a quadruple resonance. Active stabilization will be done using a piezo actuator glued to a cavity mirror. In this case, one has to take into account the piezo response curve and how it responds to small voltage fluctuations. The maximum displacement of the piezo has to allow shifting the cavity resonance by at least one FSR for the cavity modes one needs to stabilize. This is necessary to allow the cavity lock to compensate drifts at high and intermediate frequencies. 

\section{Conclusion}
Our work presents a novel design for cavity-enhanced sources of polarization-entangled photons, where the photons are created in two SPDC crystals inside a single cavity. We provide a comprehensive study of the feasibility of the design for different types of phase matching and for the cases of near-degenerate as well as highly non-degenerate wavelengths. We find that combining type-II phase matching and highly-nondegenerate wavelengths should allow the generation of high-fidelity narrowband entangled photon pairs for realistic experimental parameters. Our approach allows achieving narrow bandwidths with a simple Fabry-P\'{e}rot cavity and only moderate spectral filtering to suppress the emission of unwanted frequency modes. No spectral filtering is needed if one adds dispersive elements to the cavity. While this results in a less compact cavity design, a longer cavity offers the opportunity to achieve a significantly narrower linewidth. Our approach provides essential considerations when implementing narrow-band cavity-enhanced SPDC sources of entanglement for a variety of applications in future quantum networks. 

\acknowledgments
\v{Z}P and RK acknowledge support by the Slovenian Research Agency under contracts no. J2-2514, N1-0180, P1-0416, J1-9145, P1-0125. LU acknowledges funding by the Slovenian Research Agency under contracts no. J1-2458 and P1-0416. All authors were supported by the Republic of Slovenia (MVZI) and the European Union - NextGenerationEU (SiQUID-101091560).

\bibliography{bibliography_list}

\appendix

\section{Cluster spacing using higher-order approximations}
\label{app:higherorder}
In the main text, we used a linear approximation for calculating the cluster spacing. For most of the cases we considered there, this approximation yields the same results as a more detailed, higher-order calculation. The latter uses the definition of cluster spacing by Eckardt et al~\cite{Eckardt1991}. They found that the sum of two mode numbers
$m = m_{1} + m_{2}$ will always be very close to an integer value if one is on a double resonance. This results from the fact that the sum of the signal and idler frequencies has to be equal to the pump frequency.

The cluster spacing $\Omega$ can then be determined from an integer change in the joint mode number:
\begin{align}
\Delta m = \pm 1.
\label{eq:deltaM}
\end{align}
\begin{figure}[h]
    \centering
    \includegraphics[width=0.225\textwidth]{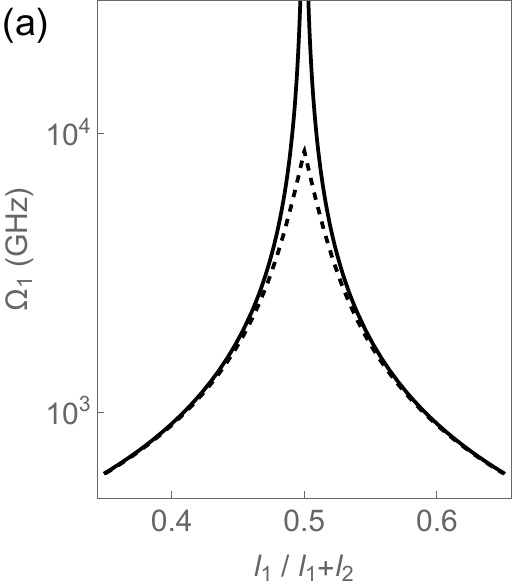} \includegraphics[width=0.235\textwidth]{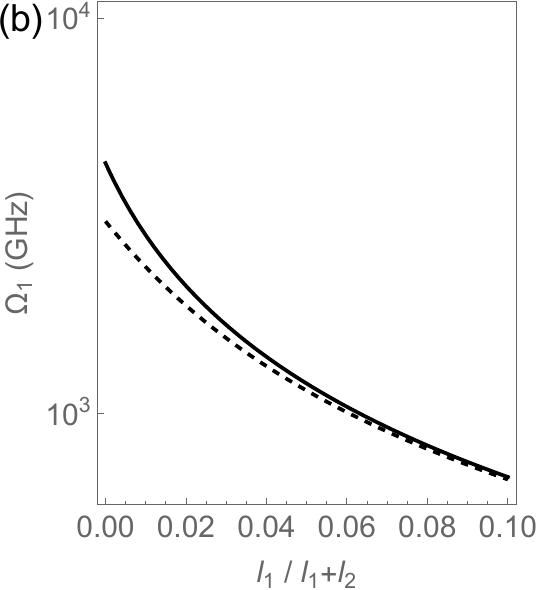}
    \\
    \includegraphics[width=0.23\textwidth]{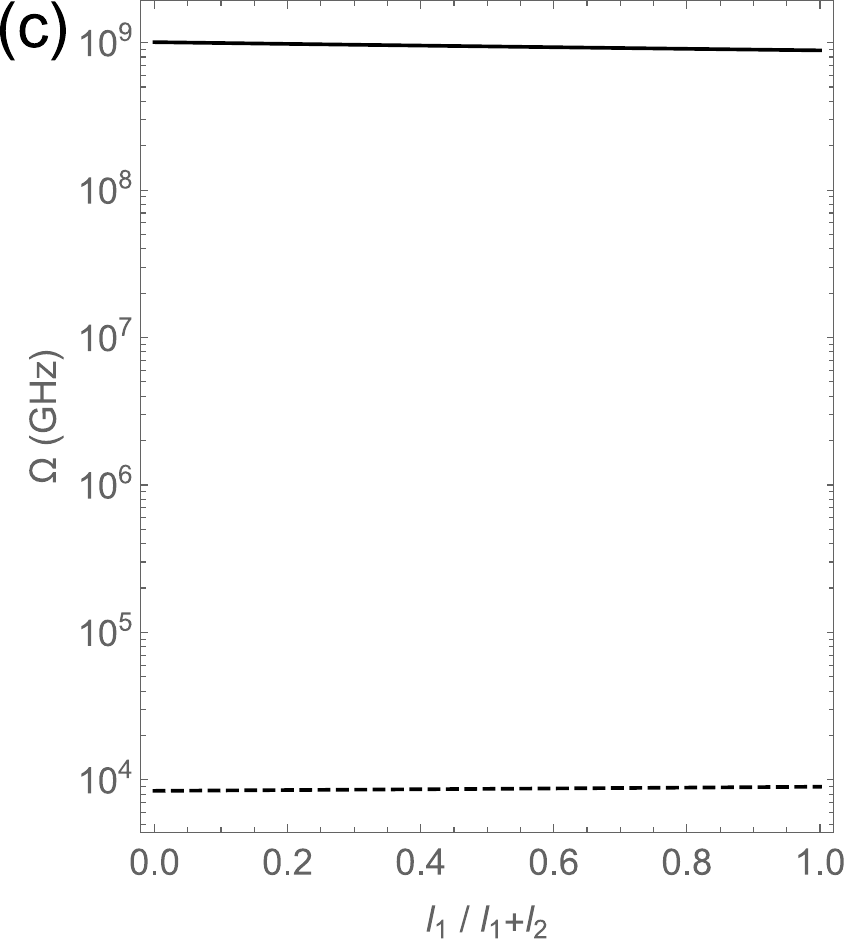} \includegraphics[width=0.23\textwidth]{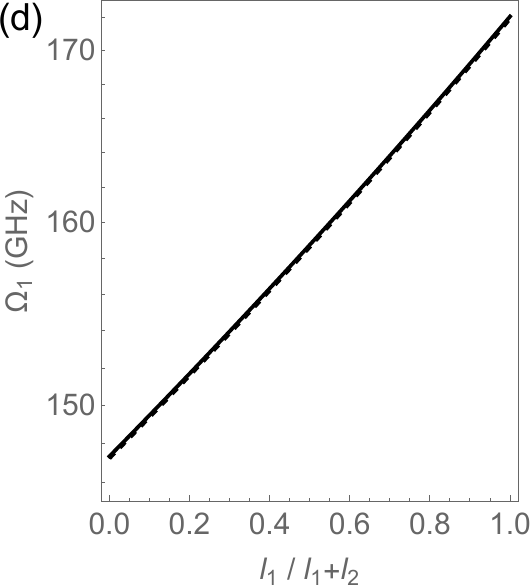}
    \caption{Comparison of the values for the cluster spacing determined using the first-order (solid) and the second-order approximation (dashed). The left-hand panels are for the near-degenerate wavelengths 
    centered around $1550\,$nm with a frequency difference of $80\,$MHz. The right-hand panels are for the strongly non-degenerate case of wavelengths $780.24\,$nm and $1550\,$nm. The top panels are for type-II phase matching (PM), the bottom ones are for type-0 PM. We assumed a joint crystal length of $l_1+l_2=10$ mm. This shows that the first and the second-order approximation only differ significantly for near-degenerate PM: either for type-II PM and near-identical crystal lengths, or for type-0 PM.}
    \label{fig:SecondOrder}
\end{figure}

Apart from the frequencies of the optical modes involved, the cluster spacing can also depend on other system parameters.
To take this into account, Eckert et al~\cite{Eckardt1991} expand the joint mode number to second order in frequency and to first order in other parameters like, e.g., the crystal temperature $T$:
\begin{equation}
    \Delta m = \left(\frac{\partial m}{\partial \omega_{j}}\right)_{\omega_{p}} \Omega +
    \frac{1}{2} \left(\frac{\partial^{2} m}{\partial \omega^{2}_{j}}\right)_{\omega_{p}} \Omega^{2} + \left(\frac{\partial m}{\partial T }\right) \left( T - T_{0} \right).
    \label{eq:modeNumberChangeTemp}
\end{equation}

Using the following relation resulting from energy conservation:
\begin{align}
\left( \frac{\partial f}{\partial \omega_s} \right)_{\omega_{p}} = \frac{\partial f}{\partial \omega_s} - \frac{\partial f}{\partial \omega_i},
\end{align}
one can calculate the partial derivatives occurring in Eq.~\ref{eq:modeNumberChangeTemp}:
\begin{align}
    \begin{split}
        \left( \frac{\partial m}{\partial \omega_s} \right)_{\omega_{p}} &= \frac{\partial m_{1}(\omega_s)}{\partial \omega_s} - \frac{\partial m_{2}(\omega_i)}{\partial \omega_i},\\
        \left( \frac{\partial^2 m}{\partial \omega_s^2} \right)_{\omega_{p}} &= \frac{\partial^2 m_{1}(\omega_s)}{\partial \omega_s^2} + \frac{\partial^2 m_{2}(\omega_i)}{\partial \omega_i^2}.
    \end{split}
    \label{eq:derivatives}
\end{align}
The cluster spacing can be calculated by solving Eq.~\ref{eq:deltaM}. This yields four possible solutions. We consider the solution yielding the lowest cluster spacing frequency because this determines the distance to the next closest cluster. In Fig.~\ref{fig:SecondOrder}, we provide a comparison of the cluster spacing using the linear approximation we used in the main text and the cluster spacing derived from integer changes of the joint mode number. One can see deviations between the solutions for the first and second-order approximations close to parameters where the cluster spacing diverges. Examples are the case of a near-degenerate source close to equal crystal lengths ($l_1\approx l_2$) and the case of non-degenerate frequencies at $l_1\to0$. 

\section{The fidelity with the target Bell state}
\label{app:fidelity}

In order to quantify the quality of the state produced by the source, we calculate its fidelity with the polarization-entangled Bell state we target. Here, we will assume $\vert\psi^{-}\rangle$. The biphotons generated will be a superposition of two-photon amplitudes of SPDC processes in crystal 1 and 2 for all possible frequency combinations of signal and idler photons within the SPDC bandwidth. These amplitudes will only couple efficiently to a cavity mode if the respective frequencies are resonant with the cavity. For type-II phase matching, we can write the two-photon state as:
\begin{equation}
\begin{split}
\vert\Psi\rangle = \sum_{j=-\infty}^{\infty}\Big(A_j^1a^\dagger_H(\omega_s+j\Omega_{1})a^\dagger_V(\omega_i-j\Omega_{1}) +\\
A_j^2a^\dagger_V(\omega_s+j\Omega_{2})a^\dagger_H(\omega_i-j\Omega_{2})\Big)\vert\text{vac}\rangle.
\end{split}
\end{equation}
Here, we assume that $\omega_s$ and $\omega_i$ both are resonant with the cavity. The amplitude for a given combination of signal and idler frequencies to the overall state is given by the normalized amplitude $A_j^\alpha=\mathrm{sinc}(l_\alpha\Delta k_{j,\alpha}/2)$ of the SPDC process at that frequency. The index $\alpha=1,2$ denotes whether the SPDC process is happening in the first or the second crystal.
The pump polarization is chosen such that the absolute values of the amplitudes for the central frequencies are equal ($\vert A_0^1\vert=\vert A_0^2\vert$), and that their relative phase matches that of the target Bell state. For example, $A_0^1=-A_0^2$ for $\vert\psi^{-}\rangle$. All of the other frequency contributions with $n\neq0$ lower the fidelity with the targeted Bell state, $F=\vert\langle\Psi\vert \psi^{\pm}\rangle\vert^2$.

\end{document}